# ON THE ECCENTRICITY OF PROXIMA b

Robert A. Brown[1,2]

[1]Space Telescope Science Institute  3700 San Martin Drive  Baltimore, MD  21218, USA, rbrown@stsci.edu

## ABSTRACT

We apply Monte Carlo projection to the radial-velocity data set that Anglada-Escudé et al. (2016) use for the discovery of Proxima b. They find an upper limit to the orbital eccentricity of $\varepsilon < 0.35$. To investigate the eccentricity issue further, we calculate a suite of mono- and bi-variate densities of $\varepsilon$. After discarding apparent artifacts at $\varepsilon \approx 0$ and $\varepsilon \approx 1$, we find that $\varepsilon$ has a tri-modal sampling distribution—three chimeras or *types* of orbit compatible with the RV data set. The three modes (peaks) in the density of $\varepsilon$ are located at $\varepsilon = \{0.25, 0.75, 0.95\}$, with relative weights $\{0.79, 0.10, 0.11\}$. Future RV observations will clarify which of the three chimeras represents the true eccentricity of Proxima b. The most-likely estimate is $\varepsilon_{est} = 0.25$, and our lower limit is $\varepsilon_{llim} = 0.025$. Our strategic, long-term goal is to elevate the orbital analyses of exoplanets to meet the challenges of sometimes complex probability density distributions.

## 1. INTRODUCTION

Recently, Anglada-Escudé et al. (2016)—hereinafter "A-E"— announced the discovery, by the radial-velocity (RV) technique, of Proxima b, an Earth-class planet, possibly in a Goldilocks orbit: not too hot and not too cold for liquid water to exist on the surface. The host star, Proxima Centauri, is an M5.5 dwarf with the distinction of being closest known star to Earth. Almost overnight, the Proxima system has become a compelling object of research related to the questions of life in the universe.

In this paper, we use a technique called "Monte Carlo (MC) projection" (Brown 2004, 2016), to study the probability densities of the RV orbital parameters of Proxima b, particularly its orbital eccentricity ($\varepsilon$), for which A-E report an upper limit, $\varepsilon < 0.35$. We show that MC projection recovers the A-E estimates of the orbital elements—all except $\varepsilon$.

The orbital eccentricities of exoplanets are of high scientific interest for a variety of reasons. First, the value of $\varepsilon$ is a clue to the formation scenario, because accretion in a massive disk tends to circularize a planetary orbit, while a thin disk may fail to do so. Second, increasing $\varepsilon$ raises the mean stellar flux incident on the planet, thereby increasing the mean surface temperature and potentially affecting habitability (Williams & Pollard 2002). Third, increasing $\varepsilon$ increases tidal heating, which tends to decrease $\varepsilon$ and warm the planet. Fourth, gravitational perturbations from additional, as-yet-undiscovered companions of Proxima—or from $\alpha$ Centauri A and/or B—could increase $\varepsilon$. Fifth, $\varepsilon$ increases the mean apparent separation of star and planet, which increases a planet's direct detectability (assuming it lies outside the central obscuration of a direct-imaging system). These issues, impacts, and implications of Proxima b's eccentricity are complicated by a further question: is the current value of ε original or has it evolved? For perspectives of the current state of play regarding the eccentricity of Proxima b, see Barnes et al. (2016) and Meadows et al. (2016).

---

[2] 1017 N. Cowboy Canyon Drive, Green Valley, AZ  85614

The immediate goal of this paper is to improve our understanding of Proxima b's orbital eccentricity by applying MC projection to the RV data set of A-E. Our strategic, long-term goal is to elevate the orbital analyses of exoplanets to meet the challenges of sometimes complex probability density distributions.

In our treatment of the Proxima b orbit, we fit seven RV orbital parameters: (1) the semi-major axis ($a$); (2) $\varepsilon$, the orbital eccentricity; (3) the mean anomaly ($M_0$) at the reference epoch $t_o$ = JD 2451634.73; (4) the stellar argument of periastron ($\omega_0$); (5) the minimum planetary mass, $m_{min} = m_p \sin i$, where $i$ is the unknown inclination angle and $m_p$ is the unknown true planetary mass; (6) the RV offset velocity ($V$); and (7) the RV acceleration ($\dot{V}$).

In the A-E treatment, the seven fitted RV orbital parameters are different from ours: (1′) the orbital period ($P$), where

$$P = \sqrt{\frac{4\pi^2 a^3}{G(m_p + m_s)}}, \qquad (1)$$

where the A-E value of the stellar mass is $m_s$ = 0.120 ± 0.015 solar masses, and $G$ is the gravitational constant; (2′) the eccentricity $\varepsilon$; (3′) the mean longitude ($\lambda_0$) at $t_o$, where

$$\lambda_0 = \omega_0 + M_0; \qquad (2)$$

(4′) stellar argument of periastron $\omega_0$; and (5′) the Doppler semi-amplitude ($K$), where

$$K = \frac{m_{min}}{\sqrt{1-\varepsilon^2} \sqrt[3]{\frac{P}{2\pi G}(m_p + m_s)^2}}; \qquad (3)$$

(6′) RV offset $V$; and (7′) RV acceleration $\dot{V}$. A-E report neither their estimated value of $V$ nor the value of $\varepsilon$ they assumed in Equation (3) when computing $m_{min}$ from $K$.

Table 1 summarizes the orbital parameters and their values as reported by A-E. Table 1 has ten rows for ten orbital parameters, where seven are produced by fits to the RV data set, and three are generated by Equations (1–3). While A-E and this paper differ in the choice of the seven fitted parameters, this difference is unimportant as long as we agree on $m_s + m_p$. This should not be an issue, because we adopt A-E's value of the stellar mass, and because both treatments ignore $m_p$ with respect to $m_s$.

Table 1 Orbital Parameters of Proxima b Based on A-E.

| Row | Quantity | Symbol | Source | A-E value | A-E 68% confidence | Units |
|---|---|---|---|---|---|---|
| 1 | semi-major axis | $a$ | eq. (1) | 0.0485 | 0.0526 / 0.0434 | AU |
| 2 | period | $P$ | fit (1′) | 11.186 | 11.187 / 11.184 | days |
| 3 | eccentricity | $\varepsilon$ | fit (2′) | < 0.35 | — / — | n/a |
| 4 | mean anomaly at $t_0$ | $M_0$ | eq. (2) | — | — / — | degree |
| 5 | mean longitude at $t_0$ | $\lambda_0$ | fit (3′) | 110 | 118 / 102 | degree |
| 6 | stellar argument of periastron | $\omega_0$ | fit (4′) | 310 | 360 / 0 | degree |
| 7 | minimum mass | $m_{min}$ | eq. (3) | 1.27 | 1.46 / 1.10 | Earth mass |
| 8 | Doppler semi-amplitude | $K$ | fit (5′) | 1.38 | 1.59 / 1.17 | m/sec |
| 9 | RV offset | $V$ | fit (6′) | — | — / — | m/sec |
| 10 | RV acceleration | $\dot{V}$ | fit (7′) | 0.086 | 0.395 / –0.223 | m/sec/yr |

Notes. Dashes indicate where the A-E paper does not report a value. A-E does not explain the upper limit on $\varepsilon$ nor how the confidence limits of $\lambda_0$ and $\omega_0$ are compatible.

MC projection involves creating a large sample of orbital solutions—100,000 solutions in this research. Each solution is produced by a least-squares fit to a unique *synthetic* data set statistically equivalent to the true data set. Our implementation of MC projection is based on the recipe in §15.6 of Press et al. (2007), and is further described, developed, and applied to RV data sets in Brown (2004, 2016).

The advantages of MC projection are twofold, at least. First, it provides a synoptic picture of the probability densities of orbital parameters, from which confidence intervals and regions are readily computed. Second, MC projection can detect the presence of multiple *types* of orbital solutions compatible with the true data set. Those features—called "chimeras"—can bias estimates of orbital parameters if they are not found and taken into account (Brown 2016, §6). Chimeras arise when the data set does not adequately constrain the *type* of orbit that is involved. Brown (2004) studied the chimeras in HD 72659 b, and Brown (2016) found chimeras in 16 of 27 RV data sets of Jupiter-class planets that are otherwise suitable for characterization by direct imaging. In this paper, we find three chimeras in the RV data set for Proxima b.

In §2, we discuss our two types of RV data sets—true and synthetic—and their associated orbital solutions, which populate the sampling distribution for this research. In §3, we study the MC projection of the RV data set of A-E. In §4, we summarize our findings.

## 2. RV DATA SETS AND ORBITAL SOLUTIONS

In our treatment, the *true* A-E data set ($\mathcal{D}_{true}$) is the entire, union data set of 214 RV measurements made available at the *Nature* website for the online version of the A-E discovery paper. This data set is the union of 144 RV measurements (RVMs) from the High Accuracy Radial Velocity Planet Searcher (HARPS) and 70 RVMs from the Ultraviolet and Visual Echelle Spectrograph (UVES). (A-E refers to two additional, as-yet unreported UVES RVMs—72 in all.)

The *true* data set has the form:

$$\mathcal{D}_{true} = \left( (t_1, u_1, \delta u_1), (t_2, u_2, \delta u_2), \ldots (t_n, u_n, \delta u_n) \right), \quad (4)$$

where $n = 144$ is the number of RVMs in the data set, $t$ is the epoch of an RVM, and $u$ and $\delta u$ are the value and uncertainty of an RVM, respectively.

A *synthetic* data set is created by a "jiggling" process:

$$\mathcal{D}_{syn} = \mathcal{J}[\mathcal{D}_{true}] = \left( (t_1, uu_1, \delta u_1), (t_2, uu_2, \delta u_2), \ldots (t_n, uu_n, \delta u_n) \right), \quad (5)$$

where $uu$ is a normal random variate with mean $u$ and standard deviation $\delta u$, and $\mathcal{J}$ is the jiggling process regarded as a function.

In our treatment, the elements of an RV orbital-solution vector ($p$) are:

$$p = (a, \varepsilon, M_0, \omega_0, m\sin i, V, \dot{V}) = \mathcal{A}[\mathcal{D}], \quad (6)$$

where $\mathcal{A}$ is the process of least-squares fitting regarded as a function of $\mathcal{D}$. Our implementation of $\mathcal{A}$ has the following features. (1) The search for the minimum of $\chi^2$ is performed in MATHEMATICA using the Levenberg–Marquardt algorithm. (2) The "guess" or starting vector ($p_0$) in each fit is a random variable, distributed uniformly between the lower and upper ranges in Table 2. (3) The RVMs are equally weighted.

To gain confidence that the search is finding a *global* minimum of $\chi^2$, we have performed an experiment, as follows. We created a sample of 10,000 synthetic data sets, fit each with 10 random starting vectors, and computed the mono-variate PDF of $\varepsilon$ (see §3.1). The result is indistinguishable from Fig. 1, indicating that the Levenberg–Marquardt algorithm is robust in this application, presumably due to a paucity of local but not global minima of $\chi^2$.

Table 2. Ranges of the Starting Points in the Searches for the Minimum of $\chi^2$.

| $i$ | parameter | lower range | upper range | units |
|---|---|---|---|---|
| 1 | $a$ | 0.0485 | | au |
| 2 | $\varepsilon$ | 0.05 | 0.95 | n/a |
| 3 | $M_0$ | 0 | 360 | degree |
| 4 | $\omega_0$ | 0 | 360 | degree |
| 5 | $m_{min}$ | 0 | 10 | Earth mass |
| 6 | $V$ | −10 | 10 | m/sec |
| 7 | $\dot{V}$ | −10 | 10 | m/sec/yr |

Note. The starting point for $a$ is not random but fixed at the A-E value in Table 1. Fig. 3 shows that the fitted values of $a$ are not biased by the starting point.

The three orbital parameters not fitted are produced by Eq. (1–3).

## 3. MONTE-CARLO PROJECTION OF THE A-E RV DATA SET

We obtain our random sample $\{p\}$ of 1,000,000 parameter vectors by solving Equation (6) for a sample of 100, 000 jiggled data sets $\{\mathcal{D}_{syn}\}$, prepared according to Equation (5). From $\{p\}$, we draw mono-variate sub-samples $\{p_i\}$ and bi-variate sub-samples $\{p_i, p_j\}$, where $i \neq j$ and $i, j = 1$–$10$. From the sub-samples, we compute empirical probability distribution functions (PDFs), using the histogrammatic method, with one- and two-dimensional bins.

*3.1 The Mono-Variate PDF of $\varepsilon$.*

Fig. 1 shows the mono-variate PDF of $\varepsilon$, with features summarized in Table 3. Fig. 2 shows the associated cumulative distribution function (CDF).

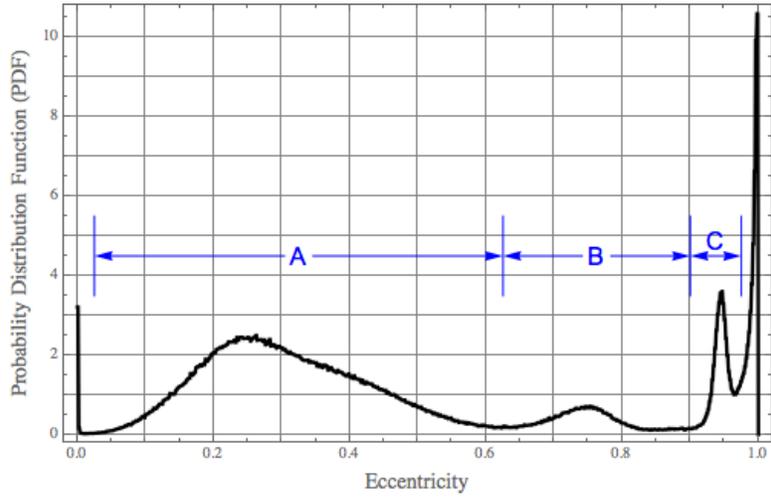

Fig. 1. The PDF of the orbital eccentricity ε of Proxima b. We regard the two end features, at ε ≈ 0 and ε ≈ 1, as artifacts of the fitting process—and ignore them.

Each range of ε—A, B, and C—hosts a chimera.

Table 3 summarizes the information in Fig. 1. Fig. 2 shows the associated CDF.

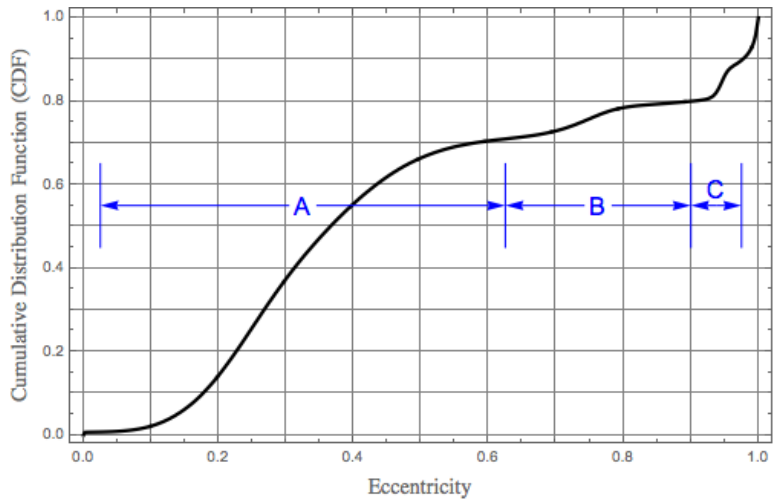

Fig. 2. The CDF of the orbital eccentricity ε of Proxima b. The PDF is shown in Fig. 1.

Table 3. The Taxonomy and Descriptive Statistics of the Eccentricity Sample {ε}

| Zone | Fraction | $\varepsilon_-$ | $\varepsilon_+$ | Mean ε | Median ε | St. dev. ε |
|---|---|---|---|---|---|---|
| $\varepsilon \approx 0$ | 0.016 | 0.000 | 0.025 | 0.0007 | 0.00003 | 0.0028 |
| A | 0.699 | 0.025 | 0.625 | 0.305 | 0.292 | 0.117 |
| B | 0.090 | 0.625 | 0.900 | 0.747 | 0.744 | 0.060 |
| C | 0.095 | 0.900 | 0.975 | 0.947 | 0.947 | 0.014 |
| $\varepsilon \approx 1$ | 0.100 | 0.975 | 0.000 | 0.992 | 0.996 | 0.007 |
| A∩B∩C | 0.884 | 0.025 | 0.975 | 0.419 | 0.338 | 0.250 |
| All | 1.000 | 0 | 1 | 0.470 | 0.362 | 0.297 |

Note. "Fraction" means the integrated PDF between $\varepsilon_-$ and $\varepsilon_+$.

The three modes (peaks A, B, C) in the PDF of ε are located at ε = {0.25, 0.75, 0.95}, and have relative weights {0.79, 0.10, 0.11}. We expect that two of these three features are false, and one is true—but which one? The most probable feature is the one in Zone A, with 79% probability. The most-likely estimate is $\varepsilon_{est} = 0.25$, and our lower limit is $\varepsilon_{llim} = 0.025$. We expect that more information—future RV measurements of Proxima—will cause one feature to thrive and grow, while the two others will fade and evaporate. Brown (2016) shows how to estimate the improvement in the accuracy of orbital parameters to be expected from future RV observations.

*3.2 Bi-Variate PDFs Involving ε.*

Figs. 3–9 show the bi-variate PDFs of ε (abscissa) paired with seven other orbital parameters: $\{a, m_{min}, \lambda_0, \omega_0, \dot{V}, M_0, V\}$, in that order. Fig. 10 shows the bi-variate PDF of ε and the reduced $\chi^2$. In all five cases where we can compare the bi-variate results of MC projection to the results of A-E (Figs. 3–7), we find qualitative agreement with the distribution of density along the ordinate (see the magenta lines).

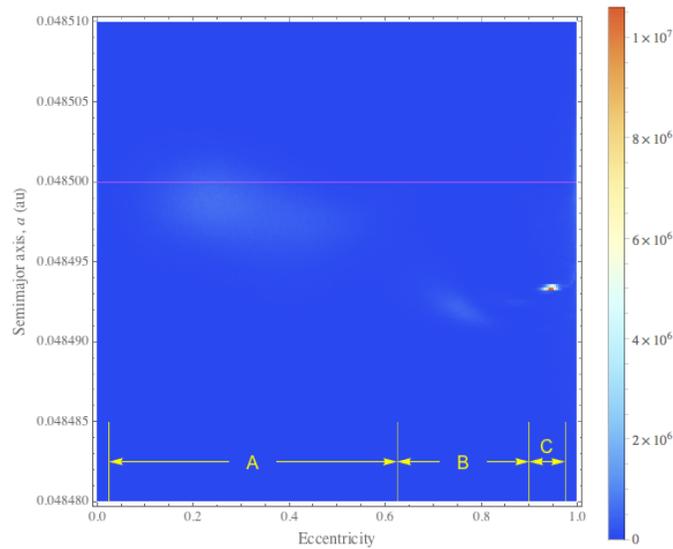

Fig 3. The bi-variate PDF of $a$ and $\varepsilon$, evidencing agreement with A-E on the period $P$, when their value of $m_s$, the stellar mass of Proxima, is assumed. Solid, magenta line: the A-E value of $a$. The A-E 85% confidence limits are outside the range of the ordinate.

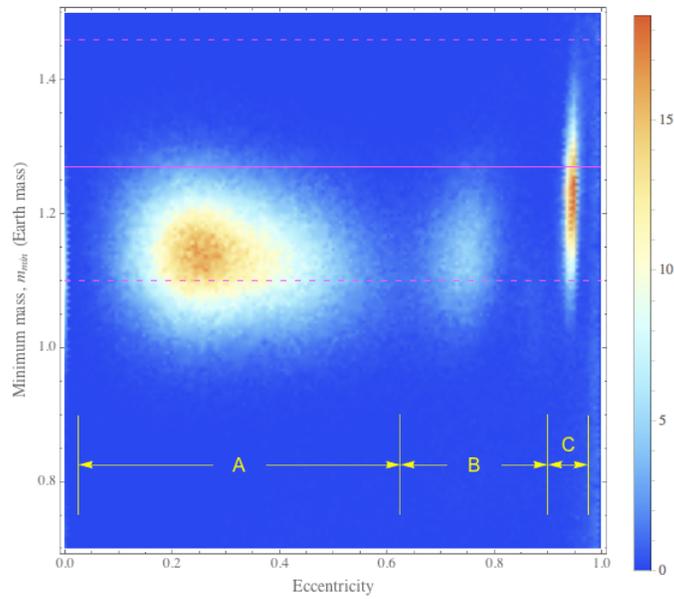

Fig 4. The bi-variate PDF of $m_{min}$ and $\varepsilon$. Solid, magenta line: the A-E value of $m_{min}$. The dashed lines: the A-E 85% confidence limits.

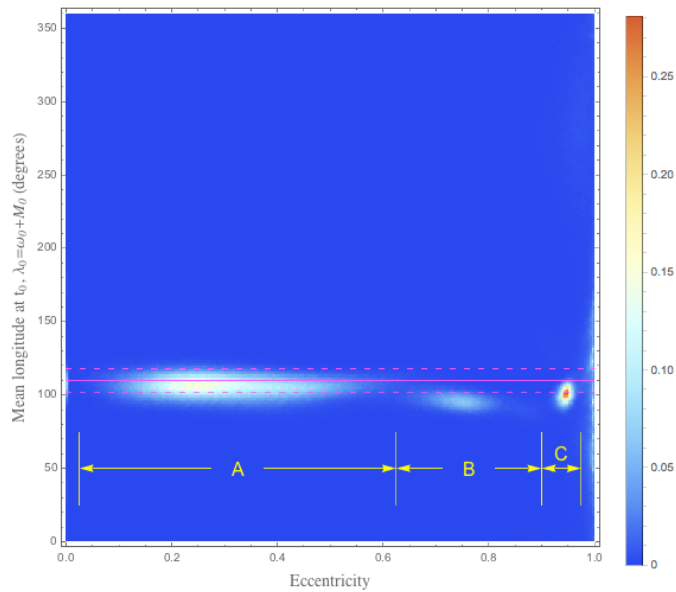

Fig 5. The bi-variate PDF of $\lambda_0$ and $\varepsilon$. Magenta lines as in the caption to Fig. 4.

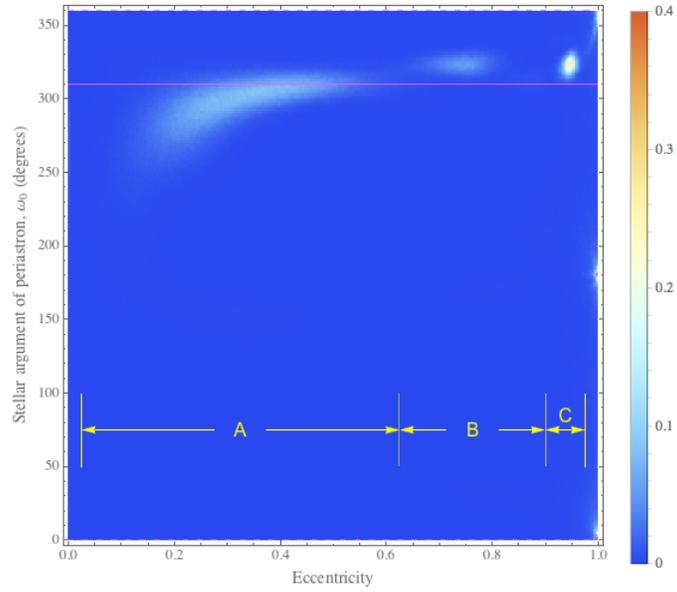

Fig 6. The bi-variate PDF of $\omega_0$ and $\varepsilon$.

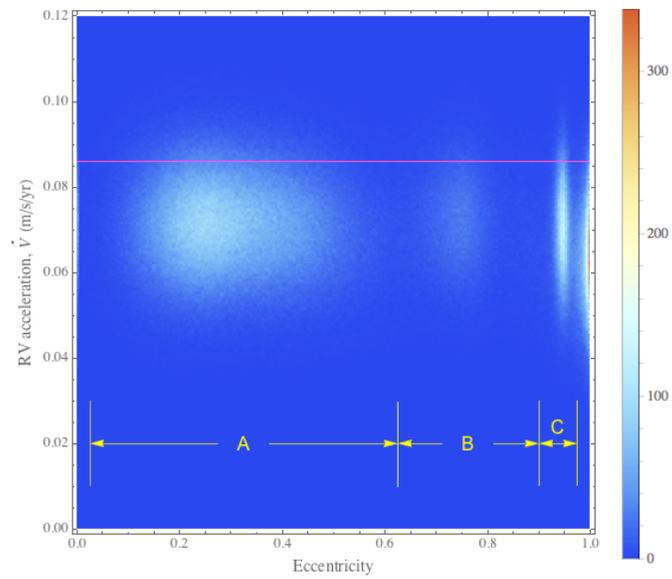

Fig 7. The bi-variate PDF of $\dot{V}$ and $\varepsilon$. Solid, magenta line: the A-E value of $\dot{V}$. The A-E 85% confidence limits are outside the range of the ordinate.

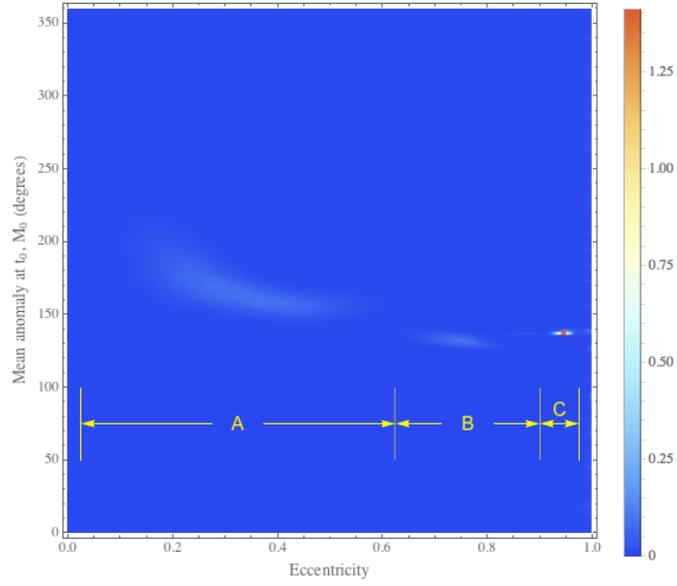

Fig 8. The bi-variate PDF of $M_0$ and $\varepsilon$. A-E do not report the value of $M_0$.

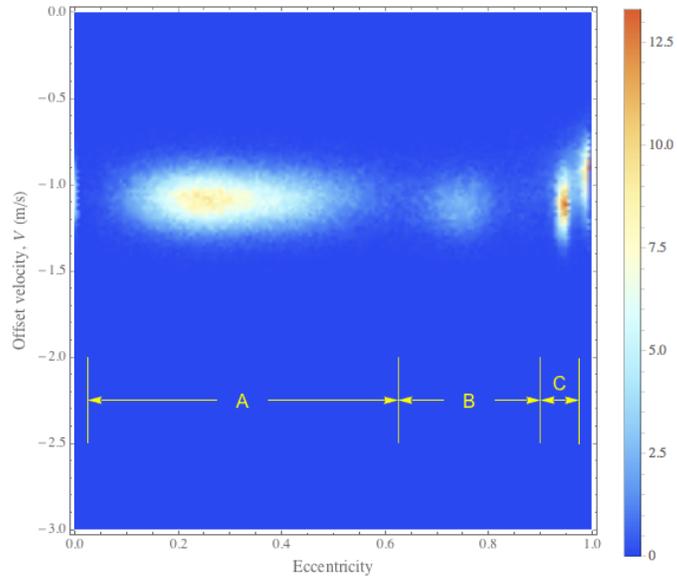

Fig. 9. The bi-variate PDF of $V$ and $\varepsilon$. A-E do not report the value of $V$.

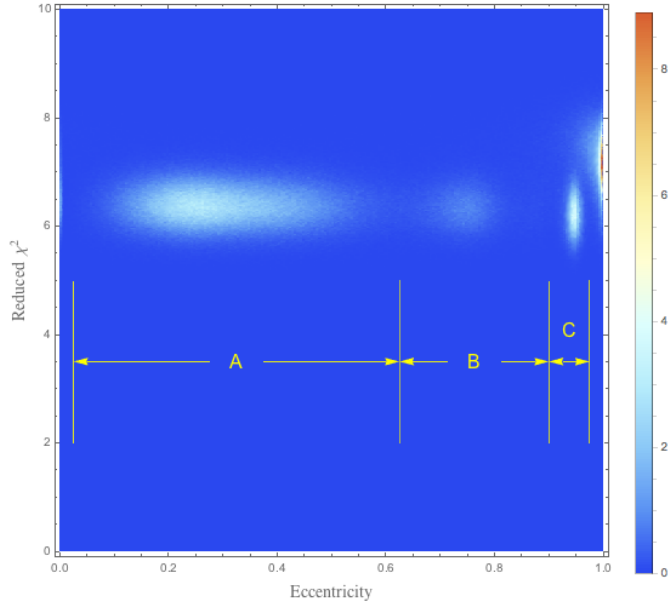

Fig.10. The bi-variate PDF of the reduced $\chi^2$ and $\varepsilon$. We expect a reduced $\chi^2$ of about unity for normally distributed measurement errors and no systematic effects.

## 4. SUMMARY

We find the current RV data set for Proxima b is compatible with three chimeras or types of orbital solutions, as evidenced by mono- and bi-variate densities involving eccentricity.

Fig. 11 shows the A-E data set, folded on itself with the measured period. Three typical solutions are superposed, one each drawn from the PDF peaks in the three zones, A to C. As the eccentricity increases, the maximum value of the RV increases sharply as $\varepsilon$ approaches unity (see the red peak). The rapidly increasing deviation between the model and the data points in the vicinity of the RV peak produces the progressive narrowing of the three density features in Fig. 1 and Figs. 3–10, as $\varepsilon$ increases in the sequence from A to C.

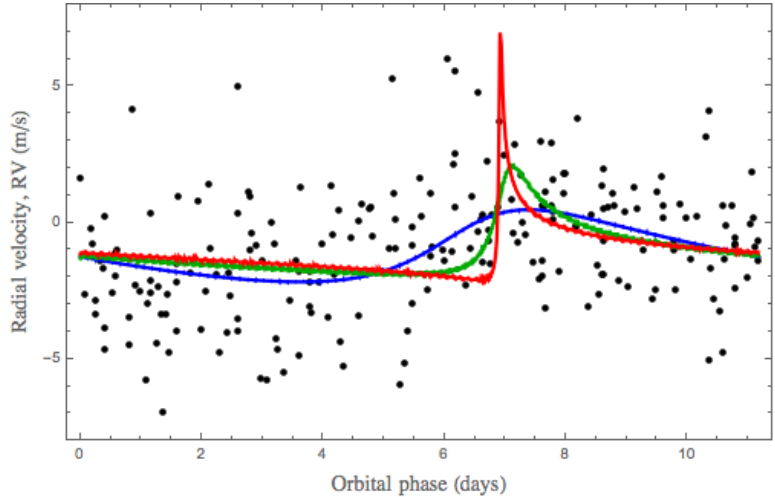

Fig. 11. The A-E data set (black) folded with the period, with typical fits in Zone A (blue), B (green), and C (red).

The three peaks in the density of $\varepsilon$ are located at $\varepsilon = \{0.25, 0.75, 0.95\}$, and have relative weights $\{0.79, 0.10, 0.11\}$. Based on the weights—and on the peak value within the highest-weighted zone, the most-likely estimate of the true eccentricity of Proxima b is $\varepsilon_{est} = 0.25$, and, by inspection, the lower limit is $\varepsilon_{llim} = 0.025$.

It is not immediately clear to the author what characteristics of the RV data set cause the bi-variate density to clump into the three zones, A, B, and C. Future RV observations will clarify which of the three chimeras represents the true eccentricity of Proxima b. Brown (2016) shows how to estimate the improvement in the accuracy of orbital parameters to be expected from future RV observations.

Our strategic, long-term goal is to elevate the orbital analyses of exoplanets to meet the challenges of the sometimes complex distributions of probability density when an RV data set offers only limited constraints on some parameters, as here for eccentricity. This paper illustrates the utility of Monte Carlo projection to gain deeper insights into the noisy RV data sets, such as those produced by Earth-type exoplanets at the limits of detectability.

MC projection could be applied to many more RV data sets, but many of those are currently off limits, in private hands, and unavailable for research by the scientific public.

We thank Stuart Shaklan for his long-time encouragement of my research, Jason Wright for the several times he has helped me locate RV data sets, Nick Schneider for his expertise and thoughtful skepticism, and Guillem Anglada-Escudé for answering some of my questions about the Proxima b discovery paper.

REFERENCES

Anglada-Escudé, G., Amado, P. J., Barnes, J., et al. 2016, Nature, 536, 438
Baluev, R. V. 2013, Astron. Comput. 2, 18
Barnes, R., Deitrick, R., Luger, R., Driscoll, P. E. et al. 2016, submitted to *Astrobiology*. http://arxiv.org/pdf/1608.06919v1.pdf
Brown, R. A. 2004, ApJ, 610, 1079
Brown, R. A. 2016, ApJ, 825:117
Cresswell, P., Dirksen, G., Kley, W., and Nelson, R. P. 2013, A & A 473, 329
Meadows, V. S., Arney, G. N., Schwieterman, E. W., Lustig-Yaeger, J., et al. 2016, submitted to *Astrobiology*. https://arxiv.org/ftp/arxiv/papers/1608/1608.08620.pdf
Press, W. B., Flannery, B. P., Teukolsky, S. A., & Vetterling, W. T. 2007, Numerical Recipes (3$^{rd}$ ed.; New York: Cambridge Univ. Press)